\documentstyle[pre,aps,tighten,multicol,epsf]{revtex}
\newcommand{\be}{\begin{equation}}
\newcommand{\ee}{\end{equation}}
\newcommand{\ba}{\begin{eqnarray}}
\newcommand{\ea}{\end{eqnarray}}

\begin{document}

%\draft

\title{ Hyperbolic Scar Patterns in Phase Space}

\author{Alejandro M. F. Rivas}

\address{Departamento de F\a'{\i}sica, Pontif\a'{\i}cia Universidade
Cat\'olica do Rio de Janeiro, \\ Caixa Postal 38071, 22452-970 Rio de
Janeiro, Brazil}

\author{Alfredo M. Ozorio de Almeida }

\address{Centro Brasileiro de Pesquisas F\'{\i}sicas\\
Rua Xavier Sigaud 150, CEP 22290-180, RJ, Rio de Janeiro, Brazil}

\date{today}
\maketitle

%twocolumn

%%%%%%%%%%%%%%%%%%%%%%%%%%%%%%%%%%%%%%%%%%%%%%%%%%%%%%%%%%%%%%%%%%%%%%%%%%%%%%
\begin{abstract}

We develop a semiclassical approximation for the spectral Wigner and Husimi functions in the neighbourhood of a classically unstable periodic orbit of chaotic two dimensional maps. The prediction of hyperbolic fringes for the Wigner function,  asymptotic to the stable and unstable manifolds, is verified computationally for a (linear) cat map, after the theory is adapted to a discrete phase space appropriate to a quantized torus. The characteristic fringe patterns can be distinguished even for quasi-energies where the fixed point is not Bohr-quantized. The corresponding Husimi function dampens these fringes with a Gaussian envelope centered on the periodic point. Even though the hyperbolic structure is then barely perceptible, more periodic points stand out due to the weakened interference.

\end{abstract}
%\begin{multicols}{2}

%\narrowtext

%%%%%%%%%%%%%%%%%%%%%%%%%%%%%%%%%%%%%%%%%%%%%%%%%%%%%%%%%%%%%%%%%%%%%%%%%%


% 71.10.-w Theories and models of many electron systems
% 71.23.-k Electronic structure of disordered solids
% 73.23.-b Mesoscopic systems

%%%%%%%%%%%%%%%%%%%%%%%%%%%%%%%%%%%%%%%%%%%%%%%%%%%%%%%%%%%%%%%%%%%%%%%%%%

%\documentstyle[preprint,aps,epsf]{revtex}
%\input tcilatex
%\topmargin 15mm
%\renewcommand{\thefigure}{\arabic{section}.\arabic{figure}}
%\input tcilatex

\section{Introduction}

The mixture of a large number of quantum states within a narrow energy window can be in principle well described by semiclassical mechanics \cite{qumec}. Yet it is not trivial to predict which classical structures will stand out within the interference pattern obtained by the various trajectories that contribute to a given quantum representation of the mixed state. So far, most attention has been directed to the immediate neighbourhood of periodic orbits. For an energy such that \cite{heller,bogomol}  the action of the orbit is nearly Bohr-quantized, there is an intensity enhancement, known as a scar.

Recently, it has been shown that a periodic orbit in a 4-D phase space actually generates a 2-D ring pattern in the Wigner function which may be more clearly distinguishable overall than the scar in its immediate  
neighbourhood \cite{fab}. These rings will be absent in a 2-D map, where the periodic orbit is reduced to a set of isolated points. However, such a mapping is the simplest case in which to study the imprints of the stable and unstable manifolds on the Wigner function of a mixed state. This effect, should also be present in the 4-D phase space, but it is orthogonal to the ring pattern and was not studdied in \cite{fab}.

The characteristic hyperbolic pattern of the mixed Wigner function in the  neighbourhood of unstable  periodic points is the main object of this paper. We shall restrict the treatment to cat maps, i.e. the quantization of linear symplectic mappings of the torus. Even though these have nongeneric spectral properties for a chaotic system \cite{hanay}, the eigenstates themselves are generically ergodic \cite{degli} in the sense of Von Neumann.
Moreover, it was shown by Keating \cite{keat2} that in this case the semiclassical theory is exact, making these maps an ideal probe into the way that the hyperbolic pattern emerges as the eigenvalue window that defines the mixed state is augmented. In the limit where we focus on a single state, the interference from many patterns washes out any visible structure, but for a sufficiently mixed state, the individual pattern is clearly discernible even when the window is not centered on a Bohr-quantized eigenvalue.

Saraceno and Voros  have already shown how the stable and unstable manifolds of the periodic points of the baker's map stamp their mark on the various iterates of the corresponding quantum propagator \cite{sarac}. For this end, they employed the Kirkwood phase space representation, which is more appropriate than the Wigner function in the case of the baker's map. As for  scars in phase space of individual states or states in a narrow energy window, most studies  have adopted the Husimi representation \cite{husimi}, which may be interpreted as a Gaussian smoothing of the Wigner function. Since the effective window that convolutes the Wigner function is usually chosen to have linear dimension $\hbar^{1/2}$, it dampens the fine structure extending along the stable and unstable manifolds that we here seek to highlight.
Indeed we show that the spectral Husimi function reproduces the fringes of the spectral Wigner function, but the amplitude is modulated by a Gaussian centred on the periodic point. Though this washes out the hyperbolic structure, the absence of long range interference allows more periodic orbits to be detected.

In section II we review Keating's definition of the spectral Wigner function for maps \cite{keat2}, introducing the Cayley parameterization of the linearized classical motion as in \cite{mcat}. We then derive the local hyperbolic form of the semiclassical contribution of the neighbourhood of a single periodic orbit to the Wigner and Husimi spectral functions. In section III we study the particular case of the cat map, following references \cite{opetor,mcat}. Not only is the basic semiclassical theory exact in this case, but our local linearized approximation is then valid throughout the torus. We can then clearly visualize the spectral Wigner function as a superposition of hyperbolic fringe patters. As the phase window defining the spectral Wigner function is decreased, more fringe patterns are superposed, but it is only as we come close to isolating the Wigner function for an individual eigenstate that the interference of the different patterns impedes their visual recognition. This interference is dampened in the Husimi spectral function so that more local scars are visible.

To conclude this introduction, we emphasize that the fringe patterns which we here analyze lie entirely within the scope of standard semiclassical theory, restricted to fairly short times. The interesting questions concerning homoclinic reccurences refer to individual states depending on the dynamics for long times \cite{heller}. Likewise, there is no need to attempt orbit ressummations \cite{resum}. These have also been presented for the cat maps \cite{catres}, even though the standard semiclassical theory of Keating \cite{keat2} is already finite in this particular case.

\section{Spectral Wigner Function}

A quantum map is characterized by its time step unitary propagator $\hat{U}
$. In terms of the 
eigenangles (or quasi-energies)  $\theta _n$  of the propagator for the eigenstates $%
|\psi _n>$,   the unitary propagator for $l$ steps is
\begin{equation}
\hat{U}^l=\sum_ne^{il\theta _n}|\psi _n><\psi _n|.
\end{equation}
The Weyl representation of $\hat{U}$ is  a function in phase space,
\begin{equation}
U^l(x)=\sum_ne^{il\theta
_n}w_n(x),  \label{ulwn}
\end{equation}
where, $x=(p,q)$ and $w_n(x)$ is the Wigner function for the
eigenstate $|\psi _n>$. We  now construct the spectral Wigner function as
\begin{equation}
W(x,\theta,\epsilon)= \sum_{l=-\infty}^{\infty} \hat{U}^l(x) e^{i l \theta}
e^{-|l| \epsilon}  \label{Wxul},
\end{equation}
so that, using (\ref{ulwn})
\begin{equation}
W(x,\theta ,\epsilon )=\sum_nw_n(x)\sum_{l=-\infty }^\infty e^{il(\theta
_n-\theta )}e^{-|l|\epsilon } . \label{Wxwn}
\end{equation}
Note that  
\begin{eqnarray}
\sum_{l=-\infty }^\infty e^{il(\theta _n-\theta )}e^{-|l|\epsilon }&\equiv &\sum_{k=-\infty }^\infty \delta _\epsilon (\theta -\theta _n+2k\pi
) =\frac{
1-e^{-2\epsilon }}{1-2e^{-\epsilon }\cos (\theta -\theta _n)+e^{-2\epsilon }}
\label{delta} ,
\end{eqnarray}
is a periodic comb of broadened delta functions. A slightly different
function was used by Keating \cite{keat2} by taking in (\ref{delta}) a sum in $%
l$ running from 0 to $\infty $. Our definition naturally generalizes
the Poisson sum formula that is recovered in the limit $\epsilon \rightarrow
0$. Alternatively, for $\epsilon >>1$,  $\delta _\epsilon (y)\approx 1$. 

Combining (\ref{Wxwn}) and (\ref{delta}), we then obtain
\begin{equation}
W(x,\theta ,\epsilon )=\sum_n w_n(x)\delta _\epsilon (\theta -\theta _n),
\label{Wxwn2}
\end{equation}
where the  sum in $k$ has been removed by taking only angles in the range 
$[0,2\pi ]$. In the limit $\epsilon \rightarrow
0$ the spectral Wigner function is a comb of delta functions on  the
eigenangles $\theta_n$, whose amplitudes are the corresponding individual Wigner functions $w_n(x)$. For values of $\epsilon $
larger than the mean level spacing, several eigenstates contribute to the
spectral Wigner function in a Lorentzian-like smoothing of energy width $\epsilon $. Evidently,  the  spectral Wigner function is only a particular representation of the density operator combining eigenstates within a narrow range of eigenangles. Thus, if $h_n(x)$ are the Husimi functions for the individual eigenstates \cite{husimi}, the spectral Husimi function corresponding to (\ref{Wxwn2}) will be 
\begin{equation}
H(x,\theta ,\epsilon )=\sum_n h_n(x)\delta _\epsilon (\theta -\theta _n).
\label{Hxwn2}
\end{equation}
Note that, contrary to the spectral Wigner function, $H(x,\theta ,\epsilon )$ is positive definite, because this property holds for each individual Husimi function and $\delta_{\epsilon}(\theta)$ is a positive function.

To obtain the semiclassical approximation for the spectral Wigner
function,   we introduce the semiclassical expression for the Weyl 
propagator, 
\be
U^l(x)= \sum_j \frac{e^{i\alpha_j}}{\left| \det (M_j^l+1)\right| ^{1/2}}{\exp }\left[ {i}\frac{{S%
}_l^j{(x)}}{{\hbar }}\right]  \label{usc}
\ee
where the sum is taken over all the classical orbits $j$ whose  center lies on the point $x$ \cite{ozrep}.  Then  $M_j^l$ stands for the monodromy matrix of the $l$'th iteration of the map and ${S}_l^j{(x)\,}$ is the center action for the orbit. In the equivalent formula for a continuous time system, the phase $\alpha =n\pi$, with $n=0$ for short times, in which case the sum reduces to a single orbit. Generally, the phases for discrete maps must be derived specifically.

Inserting (\ref{usc}) into  (\ref{Wxul}), we obtain
\begin{equation}
W_{SC}(x,\theta ,\epsilon )=  \sum_j \sum_{l=-\infty }^\infty \frac{ e^{i\alpha_j}} {\left| \det
(M_j^l+1)\right| ^{1/2}}{\exp }\left[ {i}\frac{{S}_l^j{(x)}}{{\hbar }}\right]
e^{il\theta }e^{-|l|\epsilon } . \label{Wxsc}
\end{equation}
So as to study this object in the neighbourhood of a periodic point $x_0$ of
the map, we define $X=x-x_0$ . The center action for the orbit that is close to the periodic orbit is then given by
\be
S_l^j (x)= l S_0 + X B_l X + O(X^3) , \label{sx}
\ee
where $l S_0=S_l^j(x_0)$ is the action of the periodic orbit (fixed point). For the choice of $\theta= S_0/\hbar \ mod(2\pi) $, we obtain the phase coherence of (\ref{Wxsc}) at the periodic orbit, which is said to be Bohr-quantized at this angle. $B_l$ is the symmetric matrix such that 
\begin{equation}
{\cal J} B_l = \frac{1-M_j^l}{1+M_j^l} \label{cayley},
\end{equation}
with 
\begin{equation}
{\cal J}=\left[ 
\begin{array}{c|c}
0 & -1 \\ \hline
1 & 0
\end{array}
\right] .
\end{equation}
Thus $B_l$ is the Cayley parameterization of $M_j^l$.

We can see that the term 
\begin{equation}
\left| \det (M_j^l+1)\right| ^{1/2}=2\cosh \left( \frac{\lambda l}2\right) ,
\end{equation}
where $\lambda $ is the stability exponent of the orbit. Thus  the
eigenvalues of the symplectic matrix $M_j$ are $\exp (-\lambda )$ and $\exp
(\lambda )$, corresponding to the stable and unstable directions
respectively. Using these directions as coordinate axes, we obtain 
\begin{equation}
{\cal J}B_l=\left[ 
\begin{array}{c|c}
\tanh \left( {\frac{l\lambda }2}\right)  & 0 \\ \hline
0 & -\tanh \left( {\frac{l\lambda }2}\right) 
\end{array}
\right] ,
\end{equation}
so that the contribution to the  semiclassical spectral Wigner function from the orbit that lies close to the periodic orbit takes the form
\begin{eqnarray}
W^j_{SC}(X,\theta ,\epsilon )& = &2\ Re \left\{ e^{i\alpha_j}
\left[\ {\sum_{l=0}^\infty }\frac {{\exp }\left[ {il(\frac{S_0}\hbar
-\theta )-\frac{2i}\hbar p^{\prime
}q^{\prime }\tanh \left( {\frac{l\lambda }2}\right) }\right]}{2\cosh \left( \frac{%
\lambda l}2\right) }e^{-l\epsilon }  -\frac 14  \right] \right\} \label{WSC},
\end{eqnarray}
where $X=(p^{\prime },q^{\prime })$, the  coordinates along the stable and
unstable directions respectively. This last expression shows that the phase coherence along the
stable and unstable directions also holds  along each successive hyperbola that has the stable and unstable manifolds as asymptotes. This is only a local approximation for a general nonlinear map, but it is exact for the cat map in section III. 
To perform the summation in (\ref{WSC}), we approximate ${\tanh \left( {\frac{l\lambda }2}%
\right) }=1$ and $2\cosh \left( \frac{\lambda l}2\right) =\exp \left( \frac{%
\lambda l}2\right) $  for large values of
the $\lambda l$ and $l>0$. This leads to  a geometrical series, so that
\begin{eqnarray}
W^j_{SC}(X,\theta ,\epsilon )&=&  2 \ Re \left\{ e^{i\alpha_j}  \left[ \frac{ {\exp}(-\frac{2i}\hbar p^{\prime }q^{\prime } ) }{ 1-
{\exp }\left[ i \left( \frac{S_0}{\hbar} -\theta+i(\epsilon+ 
\frac{\lambda }{2} ) \right) \right]}
   +   \sum_{l=0}^\infty  C_l (X,\theta,\epsilon) -\frac 14 \right] \right\}, \label{WSCF}
\end{eqnarray}
where the corrections, 
\be
C_l (X,\theta,\epsilon) =
{\exp }\left[ il(\frac{S_0}\hbar-\theta ) -l\epsilon  \right]
\left\{  
\frac{ {\exp}\left( -\frac{2i}\hbar p^{\prime}q^{\prime } \tanh \left( {\frac{l\lambda }2}\right)  \right) }
{ 2\cosh \left( \frac{\lambda l}2\right) } 
-  {\exp} \left( \frac{%
\lambda l}2-\frac{2i}\hbar p^{\prime }q^{\prime } \right)
 \right\},
\ee
 decrease rapidly as $l$ grows.

The dependence of the spectral Wigner function on the phase space variables is reduced to the product $ p^{\prime }q^{\prime }$ , i.e. it is constant along the hyperbolae that contain the classical trajectories. The dominant first term in the bracket of (\ref{WSCF}) specifies phase oscillations away from the asymptotes,i.e.  the stable and unstable manifolds. The amplitude of these oscillations decreases with increasing $\lambda$ and will be maximal for $\theta $ corresponding to the Bohr-quantized orbit.
We shall show in section IV that the hyperbolic structure predicted by (\ref{WSCF}) is clearly discernible in spite of the interference due to other orbits that contribute to the spectral Wigner function.

In contrast, the Husimi spectral function, 
\be
H(x,\theta ,\epsilon )=\int dx^{\prime } W(x^{\prime },\theta ,\epsilon ) w_x(x^{\prime }), \label{17}
\ee
smooths over the spectral Wigner function $W(x^{\prime },\theta ,\epsilon )$ with the Gaussian window
\be
w_x(x^{\prime })=\frac{1}{\pi\hbar}\exp{ \left\{ -\frac{\omega}{\hbar}(q^{\prime }-q)^2 - \frac{1}{\omega \hbar}(p^{\prime }-p)^2 \right\} } ,\label{18}
\ee
which is itself the Wigner function for the coherent state that has $(p,q)$ as its average momentum and position \cite{ozrep}. Generally, the axes of the ellipses in the Gaussian need not coincide with the stable and unstable directions for a periodic point. However, in this simplest case we can immediatly evaluate the integral (\ref{17}). Noticing that each term of (\ref{WSCF}) is proportional to 
\be
V_{a} (X) = e^{-2\frac{ia}{\hbar}p^{\prime }q^{\prime }}  ,
\ee
with $a$ is $1$ or $ \tanh \left( {\frac{l\lambda }2}\right)$.
We obtain each corresponding term of the spectral Husimi function as proportional to 
\be
H_{a} (X)= \left(\frac{4}{1+a^2}\right)^{\frac 12} 
\exp{\left[-
 \frac{\omega}{\hbar}\frac{a^2}{1+a^2} q^{\prime 2}-
	    \frac{1}{\omega\hbar}\frac{a^2}{1+a^2} p^{\prime 2} \right]}
Re\left\{  \exp {\left[
 \frac{-i}{\hbar}\frac{a}{1+a^2} q^{\prime }p^{\prime }
\right] }\right\} ,
\ee
where $a =1 $ for the dominant resummed term. 
Thus, in this case, the Husimi spectral function basically dampens the hyperbolic Wigner  pattern with a Gaussian amplitude modulation centered on each periodic point. The presence of the stable and unstable directions is suggested, but the scar is localized in the neighbourhood of each periodic point. Qualitatively, we expect the same behaviour in the general case where the axes of the hyperbolae do not coincide with the $(p^{\prime },q^{\prime })$ coordinates that define the coherent state basis. Though the interference fringes still exist, they will be severely dampened far from the periodic orbit. This may allow for the detection of a greater number of periodic orbit scars.

\section{The Cat Map}

We now apply the present theory to the  cat map i.e. the linear automorphism on the $2$-torus generated by
the $2\times 2$ symplectic matrix ${\cal M}$, that takes a point $x_{-}$ to a point $x_{+}$ : $x_{+}={\cal M}x_{-}\quad \mbox{mod(1)} $. In other words, there exists an integer $2$-dimensional vector ${\bf m}$ such that 
$
x_{+}={\cal M}x_{-}-{\bf m}  $. Equivalently, the map can also be studied in terms of the center generating function \cite{mcat}.
This is defined in terms of  center points 
\begin{equation}
x\equiv \frac{x_{+}+x_{-}}2  \label{xdef}
\end{equation}
and chords 
\begin{equation}
\xi \equiv x_{+}-x_{-}=-{\cal J}\frac{\partial S(x,{\bf m})}{\partial x},
\label{corgenfu}
 \label{cordef}
\end{equation}
where 
\begin{eqnarray}
S(x,{\bf m}) &=&xBx+x(B-{\cal J)}{\bf m}+\frac 14{\bf m}(B+\widetilde{%
{\cal J}}){\bf m}  \label{sx2} 
\end{eqnarray}
is the center generating function. Here $B$ is  a symmetric matrix (the Cayley
parameterization of ${\cal M}$, as in (\ref{cayley})),
while 
\begin{equation}
\widetilde{{\cal J}}=\left[ 
\begin{array}{c|c}
0 & 1 \\ \hline
1 & 0
\end{array}
\right] .
\end{equation}
We will study here the cat map with the symplectic matrix
\begin{equation}
{\cal M}=\left[ 
\begin{array}{cc}
2 & 3 \\ 
1 & 2
\end{array}
\right] \mbox{, and symmetric matrix } B=\left[ 
\begin{array}{cc}
-1 & 0 \\ 
0 & \frac{1}{3}
\end{array}
\right] . \label{mhb}
\end{equation}
This map is known to be chaotic, (ergodic and mixing) as all its periodic orbits are hyperbolic.
 The periodic
points $x_l$ of integer period $l$ are labeled by the winding numbers ${\bf %
m,}$ so that 
\begin{equation}
x_l=\left( 
\begin{array}{l}
{p_l} \\ 
{q_l}
\end{array}
\right) =({\cal M}^l-1)^{-1}{\bf m} .\label{xfix}
\end{equation}
The first periodic points of the map are the 
fixed points at $(0,0)$ and $(\frac 12, \frac12)$ and the periodic orbits of period 2 are  $[(0,\frac12)$ , $ (\frac12,0)]$, $[(\frac12,\frac16)$ , $ (\frac12,\frac56)]$, $[(0,\frac16)$ , $ (\frac12,\frac26)]$,  $[(0,\frac56)$, $(\frac12,\frac46)]$ and $[(0,\frac26), (0,\frac46)]$. The eigenvalues  of  $ {\cal M}$ are  $ e^{-\lambda} $ and $e^{\lambda}$ with $ \lambda = \ln (2 + \sqrt{3}) \approx 1.317 $. This is then the stability exponent for the fixed points, whereas the exponents must be doubled for orbits of period 2. All the  eigenvectors have directions  $v_1=(-\sqrt{3},1)$ and $v_2=(\sqrt{3},1)$ corresponding to the stable and unstable directions respectively. If we neglect the corrections in (\ref{WSCF}), the semiclassical spectral Wigner function has identical fringe patterns for all periodic orbits, except for their overall phase and the amplitude of the oscillations, where we must substitute $\lambda$ by $ n \lambda$ for an orbit of period $n$.

Quantum mechanics on the torus, implies a finite Hilbert space of dimension $ N=\frac{1}{2\pi \hbar }$, and that positions and momenta are defined to have discrete values in  a lattice of separation $\frac{1}{N}$ \cite{hanay,opetor}. 
The cat map was originally quantized by Hannay and Berry \cite{hanay} in the coordinate representation:
\be
<{\bf q}_{k}|\hat{\bf U}_{\cal M} |{\bf q}_{j}> = \left( \frac{i}{N} \right) ^{\frac12}
{\exp} \left[ \frac {i2\pi}{N}( k^{2} -j k + j^{2})\right] ,\label{uqq}
\ee
where the states $<q|{\bf q}_{j}>$ are periodic combs of Dirac delta distributions at positions $q=j/N mod(1)$, with $j$ integer in $[0,N-1]$. In the Weyl representation \cite{opetor},  the quantum map has been obtained  in \cite{mcat} as
\begin{eqnarray}
{\bf U}_{{\cal M}}(x) &=&\frac{2}{\left| \det ({\cal M} +1)\right|^{\frac 12}}  \sum_{{\bf m}}e^{i2\pi N\left[ S(x,{\bf m})\right] } \nonumber \\
&=&\frac{2}{\left| \det ({\cal M} +1)\right|^{\frac 12}}\sum_{{\bf m }}e^{i2\pi N\left[
xBx+x(B-{\cal J}){\bf m}+\frac 14{\bf m}(B+\widetilde{{\cal J}}){\bf m}%
\right] },  \label{ugxp}
\end{eqnarray}
where the center points are represented by $x=(\frac{a}{N},\frac{b}{N})$ where $a $ and $b$ are integers in $[0,N-1]$ for odd values of $N$ \cite{opetor}.
There exists an alternative definition of the torus Wigner function which also holds for even $N$. However, it is constructed on the quarter torus and this compactification scrambles the hyperbolic patterns as discussed in the Appendix.

The fact that the ${\cal M}$ matrix has equal diagonal  elements implies that the  $ B $ matrix has no off-diagonal elements. This property will be valid for all the powers of the map and, using (\ref{ugxp}), we can see that it implies in the quantal  symmetry
\be
{\bf U}_{{\cal M}}^{l}(p,q)= \left({\bf U}_{{\cal M}}^{l}(-p,q)\right)^*  = \left({\bf U}_{{\cal M}}^{l}(p,-q)\right)^* \label{qsym}.
\ee

It has been shown \cite{hanay} that the unitary propagator is periodic (nilpotent)  in the sense that, for any value of $N$ there is an integer $k(N)$ such that $\hat{\bf U}_{\cal M}^{k(N)}=e^{i\phi} $. Hence the  eigenvalues of the map lie on the $k(N)$ possible sites 
\be
\left\{ {\exp} \left[ \frac{1(2m\pi +\phi}{k(N)} \right] \right\} ,\quad 1\le m\le k(N) . 
\ee
For $k(N)< N$ there are degeneracies and the spectrum does not behave as expected for chaotic quantum systems.
In spite of the peculiarities in this spectra of quantum cat maps, it is likely that non-degenerate states are typical of chaotic maps, such as very weak nonlinear perturbations of cat maps that are known to have nondegenerate spectra \cite{matos}. Eckhardt \cite{Eckhardt}  has argued that typically the eigenfunctions of cat maps are random.

 Keating \cite{keat2} has shown   that the periodicity of the unitary propagator leads to
\begin{equation}
\sum_{m=1}^{N} w_m(x) \delta_{\theta_m , \theta_n} = \frac{1}{k(N)}\sum_{l=0}^{k(N)} \hat{ \bf U}^l(x) e^{i l \theta_n},\quad \mbox{ here }\quad \delta_{\theta_m , \theta_n}= 
\left\{
\begin{array}{ll}
1 &  \mbox{if } \theta _m = \theta _n \\
0 &  \mbox{otherwise}
\end{array}
\right. .
\label{Wxuc}
\end{equation}
The symmetry property (\ref{qsym}) conjugated with  (\ref{Wxuc}) is transmitted to  the Wigner functions for the eigenstates: 
\be
w_{n}(p,q)=w_{n}(p,-q)=w_{n}(-p,q).
\ee
The Husimi representation on the torus depends on the definition of the periodic coherent state \cite{nonen,sarac2}, with $<p>=P$ and $<q>=Q $:
\be
<P,Q|{\bf q}_k >= \sum_{j=-\infty}^{\infty} 
exp{ \left\{ 2\pi N \left[ -\frac{(j+Q-k/N)^2}{2 \omega ^2} 
-iP(j+ Q -k/N)
\right] \right\}   } .\label{35}
\ee
Then the Husimi function for the state $|\psi_n >$ is defined as  
\ba
h_n(x) & = & Tr (|\psi_n ><\psi_n |P,Q ><P,Q| )\nonumber \\ 
  &  = &   \frac{1}{N}\sum_{x_1} W_n(x_1)W_{(P,Q)}(x_1), \label{30}
\ea
where 
\be
W_{(P,Q)}(x)= \sum_{k=0}^{N-1}<{\bf q}_{2b-k}|P,Q><P,Q|{\bf q}_{k}>e^{-i\frac{2\pi }N2(b-k)a} \label{wtrans}
\ee
is the Wigner function for the coherent state. This has been studdied numerically by  Bianucci et al. \cite{sarac2} in the case of even $N$. In the appendix we show that
\ba
W_{(P,Q)}(x)& = & \sum_{j=-\infty}^{\infty}\sum_{k= -\infty}^{\infty}
\frac{N\omega^2}{2} (-1)^{2ja+2kb+jkN}  \nonumber \\
&  \times& \exp{ \left\{  2\pi N  \left[ - \omega^2 \left(P-\frac{a}{N} -  \frac{k}{2}  \right) ^2 
-\frac{1}{\omega^2}\left(Q-\frac{b}{N} -  \frac{j}{2}  \right) ^2
\right] \right\} },
\ea
which can be separated into a term that is a periodic Gaussian in phase space centered on the point $X= (P,Q)$ whereas the other terms represent Gaussians  centered on  $x= (X+\frac{{\bf m}}{2})$ . 
By definiting the Wigner function on the integer lattice ($a$ and $b$ are integer) none of the four Gaussians on the torus are oscillatory, contrary to the case in \cite{sarac2}. Thus the average over the Wigner function combines the region near $X$ with those near $(X+\frac{{\bf m}}{2})$. Fortunately, the cat map that we chose has periodic points of identical structure with this same displacement. Otherwise the hyperbolic structure would be  erased even more in the Husimi function.

Figure 1 shows the spectral Wigner function for a cat map with $N=223$. In  figure 1(a) we see the single Wigner function for the pure state with  $k=113$. For the figures 1 (b) (c) and (d) we show the spectral Wigner function $W(x,\theta ,\epsilon ) $ for the angle $\theta$ corresponding to the state with  $k=113$ but for different quasi-energy windows with respectively $\epsilon= 4\Delta, 20\Delta$ and $ 40\Delta$, where $\Delta=2\pi /N$ is the mean level spacing of the system.
These figures show clearly the imprints of the stable and unstable manifolds that are plotted for clarity in figure 1(d). 
For the two largest values of $\epsilon$ we can clearly see scars of the fixed points and also of the periodic orbit $[(0,\frac12)$ , $ (\frac12,0)]$ while for the case of smaller  $\epsilon= 4\Delta$ we can also appreciate scars of the periodic orbit,  $[(\frac12,\frac16)$ , $ (\frac12,\frac56)]$.

In figure 2  we compare the exact spectral Wigner function for a cat map with $N=223$ in  (a)  with the semiclassical approximation $W^j_{SC}(x,\theta ,\epsilon )$, correspondingly for  $\hbar= 1/(2\pi N) $, in (d), near the fixed point at $(1/2,1/2)$ whose  action is $ S_0=\frac{3}{4} $. Figure 2.(b) and 2.(c)  show respectively an horizontal and and vertical sections of the objects plotted in figures 2.(a) and 2.(b). As we can see, the agreement between the exact dynamical system and the semiclassical approximation is very good.
The modulation arrising in the exact  spectral Wigner function is due to the interference with other periodic orbits.
Figure 3 corresponds to the Wigner and Husimi spectral functions with the same energy windows. By erasing the long range extension of the hyperbolic fringes, it becomes possible to detect period-2 points of the period-2 orbit, which are not discernible in the Wigner function.

\section{Discussion}

It has become  established lore that the Husimi representation is the ideal instrument to uncover the underlying classical phase space structure coded into a bunch of closely associated quantum eigenstates. However, the Gaussian smearing of the Wigner function that generates the Husimi function washes out the characteristic hyperbolic patterns that stand out in our computations for the cat maps. Indeed the half-width of the gaussian in our figures corresponds to $\sqrt{N}\approx 15 $ points of the Wigner lattice.
The imprint of the classical hyperbolicity of the Wigner function is so clear that we can detect it even for quasi-energies that do not correspond to a Bohr-quantized periodic orbit. 

The general features exhibited by our calculations should also be discernible for nonlinear systems as, indeed, our deduction was not restricted to cat maps. 
Though the theory in section II is only local, we conjecture that distorted hyperbolae asymptotic to curved stable and unstable  manifolds  will bear fringes reaching out from the periodic point. In the case of a chaotic Hamiltonian for two degrees of freedom, this pattern should emerge in two-dimensional sections cutting the periodic orbit at a point. This plane should be transverse to that of the orbit, where the Wigner function exhibits the pattern of concentric rings discussed in \cite{fab}. Thus the full picture of the spectral Wigner function emerges as a sequence of rings in the plane of the orbit composed with the hyperbolic fringes receeding from the stable and unstable manifolds.

It is remarkable that in the case of the cat map, the hyperbolic fringes surrounding the fixed point emerge, even for quasi-energy smoothing $\epsilon$ of a few level spacings. Actually, the exponencial cut-off with time in our definition of the spectral Wigner function affords equal treatment to all periodic orbits in the denominator of (\ref{WSCF}). The reason why only one hyperbolic fringe system is visible is that for an orbit of period $n$ the lyapunov exponent is multiplied by $n$ i.e. $\lambda \rightarrow n \lambda$. As $\epsilon$ increases, the interference from the longer orbits is further diminished, revealing identical hyperbolic patterns along period-2 orbits, as predicted by our semiclassical formula. Of course, increasing $\epsilon$ further eventually washes out all the structure. In this respect, the Husimi dampening of Wigner oscillations allows for the visibility 
of a greater number of local scars.

%%%%%%%%%%%%%%%%%%%%%%%%%%%%%%%%%%%%%%%%%%%%%%%%%%%%%%%%%%%%%%%%%%%%%%%%%%

\section{Acknowledgements}

We are grateful to M. Saraceno for stimulating discussions and for providing us the coherent state on the torus together with the Wigner function pictures of this coherent states. We also aknowledge R.O. Vallejos for  his critical reading of the manuscript. We  thanks the Brazilian
agencies CNPq, FAPERJ, and PRONEX for financial support.

%%%%%%%%%%%%%%%%%%%%%%%%%%%%%%%%%%%%%%%%%%%%%%%%%%%%%%%%%%%%%%%%%%%%%%%%%%

\section*{Appendix: Wigner function for a torus coherent state}

To evaluate a discrete torus Wigner function at $x=(a/N, b/N)$ where $a$ and $b$ are half integers, we introduce the coherent state  wave function (\ref{35}) in the Wigner transform (\ref{wtrans}) to obtain 
\ba
W_{(P,Q)}(x) & = & \sum_{n=0}^{N-1}\sum_{j,j' = -\infty}^{\infty}
\exp{ \left\{  2\pi N  \left[ 
iP(j-j')
+i\frac{2(b-n) }{N}\left(P-\frac{a}{N}\right)
\right. \right.    } \nonumber \\ 
& - & \left. \left. \frac{1}{ 2\omega^2 } \left( \left( j' -Q - \frac{n}{N}\right)^2  + \left(j+Q-\frac{2b-n}{N}\right)^2 \right)
\right] \right\}    .
\ea
The substitution $j'=j'' - j $ now leads to 
\ba
W_{(P,Q)}(x)& = & \sum_{n=0}^{N-1}\sum_{j,j'' = -\infty}^{\infty}
\exp{ \left\{  2\pi N  \left[
  - \frac{1}{\omega^2 }\left(  \left(Q - \frac{b+(n+Nj)}{N}\right)^2    \right. \right. \right.  }   \nonumber \\
& + & \left. \left(\frac{b}{N}\right)^2 -  2\frac{Q}{N}(n-Nj) +  j''\left(Q-\frac{n+Nj}{N}\right) +\left(\frac{j''}{2}\right)^{2} \right)  \nonumber \\
 & + & \left. \left. i \left( \frac{2b}{N}\left(P-\frac{a}{N}\right)-\frac{2}{N}\left(P-\frac{a}{N}\right)(n+Nj) + j''P 
\right)
\right] \right\}    ,
\ea
so that we can now combine  the sum over $n$ and $j$ into a single sum over $m=n+Nj$ 
\be
W_{(P,Q)}(x)= \sum_{j=-\infty}^{\infty}\sum_{m = -\infty}^{\infty}
\exp{ \left\{  2\pi N  \left[ \frac{-1}{\omega^2}\left(\frac{m}{N}\right)^2 
- B_j\frac{m}{N} + C_j \right] \right\} },
\ee
where we also take $j''\rightarrow j  $ and 
\be
B_j = 4 \pi \left[  \frac{1}{\omega^2} \left( \frac{b}{N}-\frac{j}{2}\right) 
-  i \left(P-\frac{a}{N}\right)
\right]
\ee
and
\be
C_j =\frac{1}{\omega^2} \left[ 
Q^2 + 2 \left(\frac{b}{N}\right)^2 - Q \frac{b}{N}+2 j Q+j^2  
\right] -i \left[2 \frac{b}{N}\left(P-\frac{a}{N}\right) + j P  \right] .
\ee
We now take the Poisson transform  of the sum over $m$ so that 
\ba
W_{(P,Q)}(x) & = & \sum_{j=-\infty}^{\infty}\sum_{k= -\infty}^{\infty}
\int_{-\infty}^{\infty} dx 
\exp{ \left\{  2\pi N  \left[\frac{-1}{\omega^2}x^2 +\left(B_j +\frac{k}{N} \right) x-
C_j 
\right] \right\} } \nonumber 
\\
 & = & \sum_{j=-\infty}^{\infty}\sum_{k= -\infty}^{\infty}
\frac{N\omega^2}{2} (-1)^{2ja+2kb+jkN}  \nonumber \\
&  \times& \exp{ \left\{  2\pi N  \left[ - \omega^2 \left(P-\frac{a}{N} -  \frac{k}{2}  \right) ^2 
-\frac{1}{\omega^2}\left(Q-\frac{b}{N} -  \frac{j}{2}  \right) ^2
\right] \right\} }. \label{wcs}
\ea
Thus we obtain the superposition of four periodic gaussians, one centered at $(P,Q)$ and the other three displaced by half integer vectors as described by Bianucci et al \cite{sarac2}. However,  it is now necessary to distinguish alternative realizations of the Wigner function. In the case of even $N$, as used in \cite{sarac2}, the Wigner function is defined on the quarter torus \cite{opetor}, so this is the range of the half integer vectors $x_1$ in the definition of the Husimi function (\ref{30}). 
By centering the quarter torus near the point $(P,Q)$ the Husimi function (\ref{30}) is then dominated by the non-oscillatory Gaussian, whereas it is only the tails of the other three Gaussians that penetrates in the chosen quarter.

We have here followed the alternative course of working with odd $N$ which enables us to define the Wigner function on an integer lattice on the whole torus. This leads to intelligible scar patterns that would be folded into the quarter torus by the rule \cite{opetor}
\be
W(x+\frac{(k,j)}{2}) = (-1)^{2ja+2kb+jkN}W(x) ,
\label{w--}
\ee
which would spoil phase coherence along the hyperbolae.

In the case of the Wigner function for the coherent state, the phase factor in (\ref{w--}), which coincides with that in (\ref{wcs}), leads to four non-oscillatory Gaussians, supported by the integer lattice. The Gaussians centered on $(P,Q)$, $(P+\frac{1}{2},Q)$ and $(P,Q+\frac{1}{2})$ are positive, whereas $(P+\frac{1}{2},Q+\frac{1}{2})$ centres a negative Gaussian. All four Gaussians must then be taken into account in evaluating (\ref{30}) and it is only the periodic repetitions of these Gaussians whose tails can be neglected.

\begin{figure}[h]
\setlength{\unitlength}{1cm}
%\begin{picture}(0,6)(0,0)
%\put(0.1,0){\epsfxsize=8cm \epsfbox[83 290 512 552]{fig1.ps} }
%\centerline {\epsfxsize=17cm \epsffile{fig1.ps} }
\begin{picture}(0,16)(0,0)
\put(0.0,8.0){\epsfxsize=8cm\epsfbox[0 0 550 550]{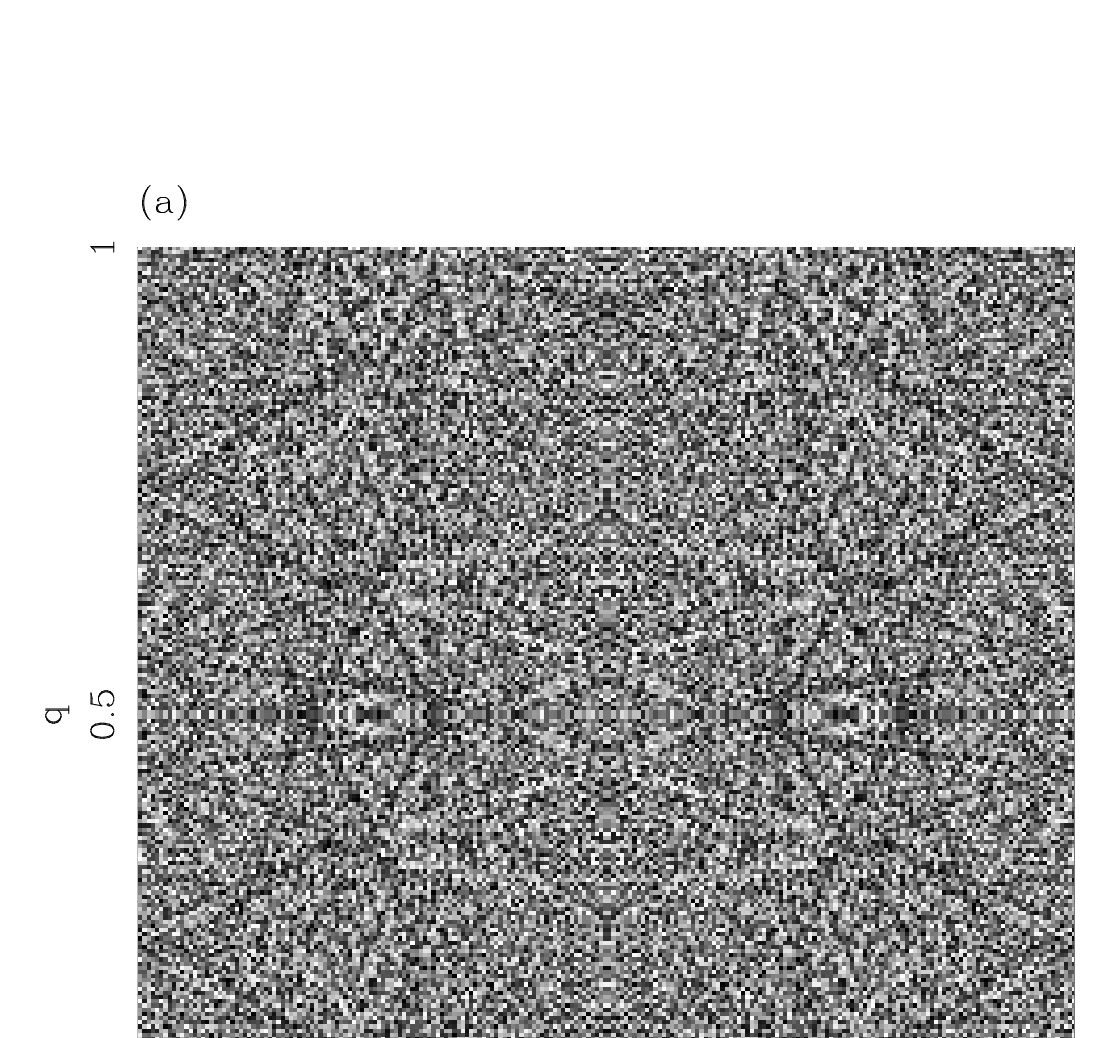}}
\put(8.0,8.0){\epsfxsize=8cm\epsfbox[0 0 550 550]{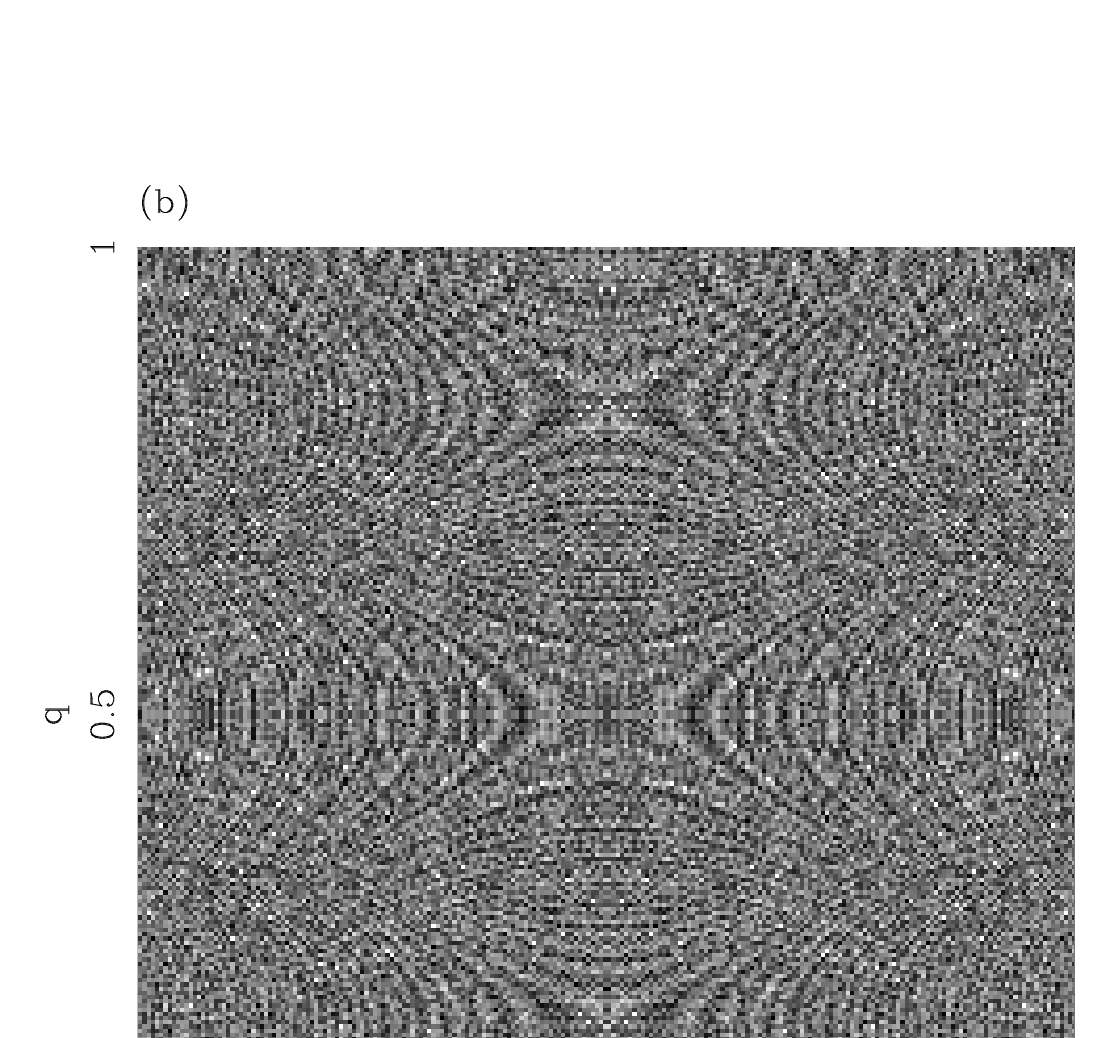}}
\put(0.0,0.0){\epsfxsize=8cm\epsfbox[0 0 550 550]{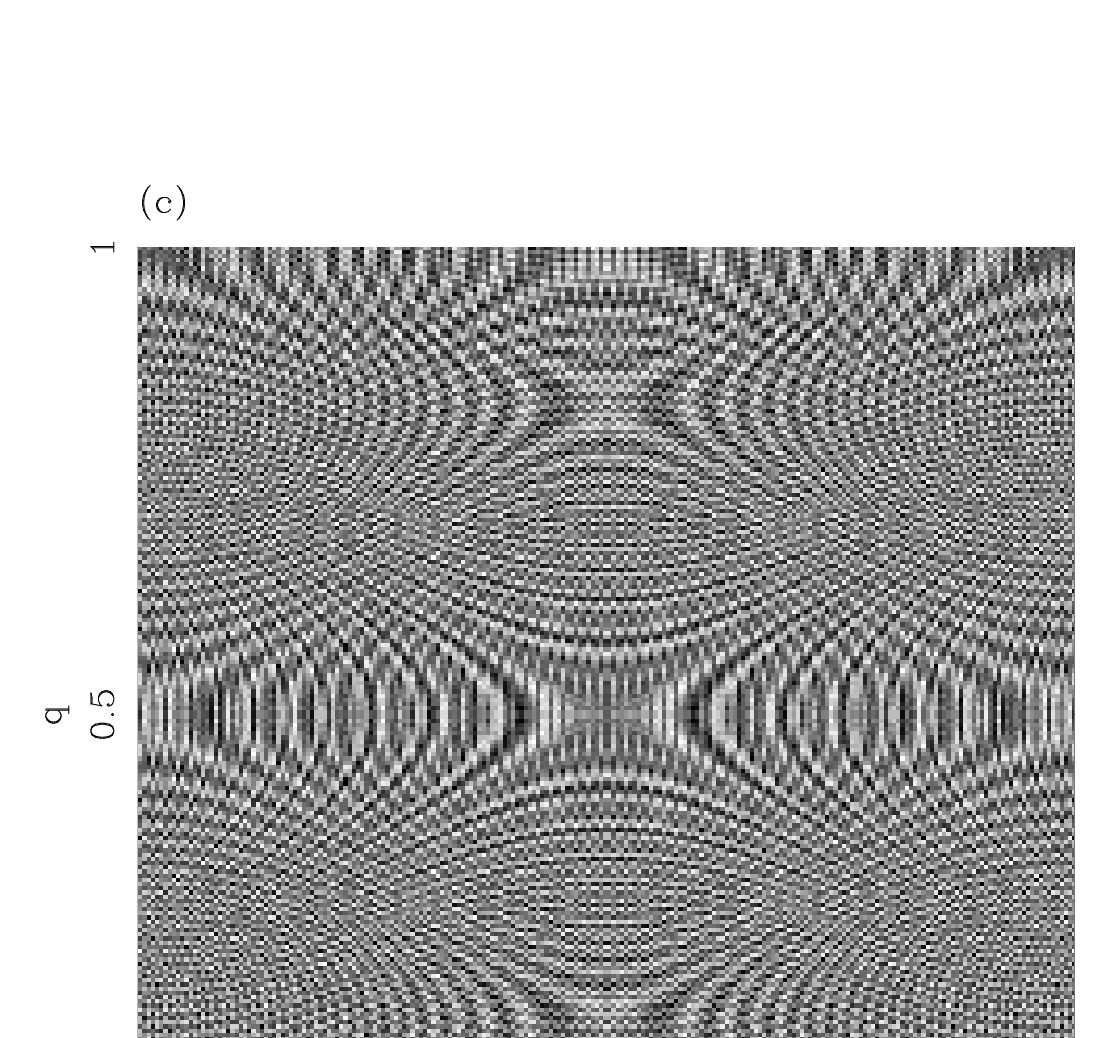}}
\put(8.0,0.0){\epsfxsize=8cm\epsfbox[0 0 550 550]{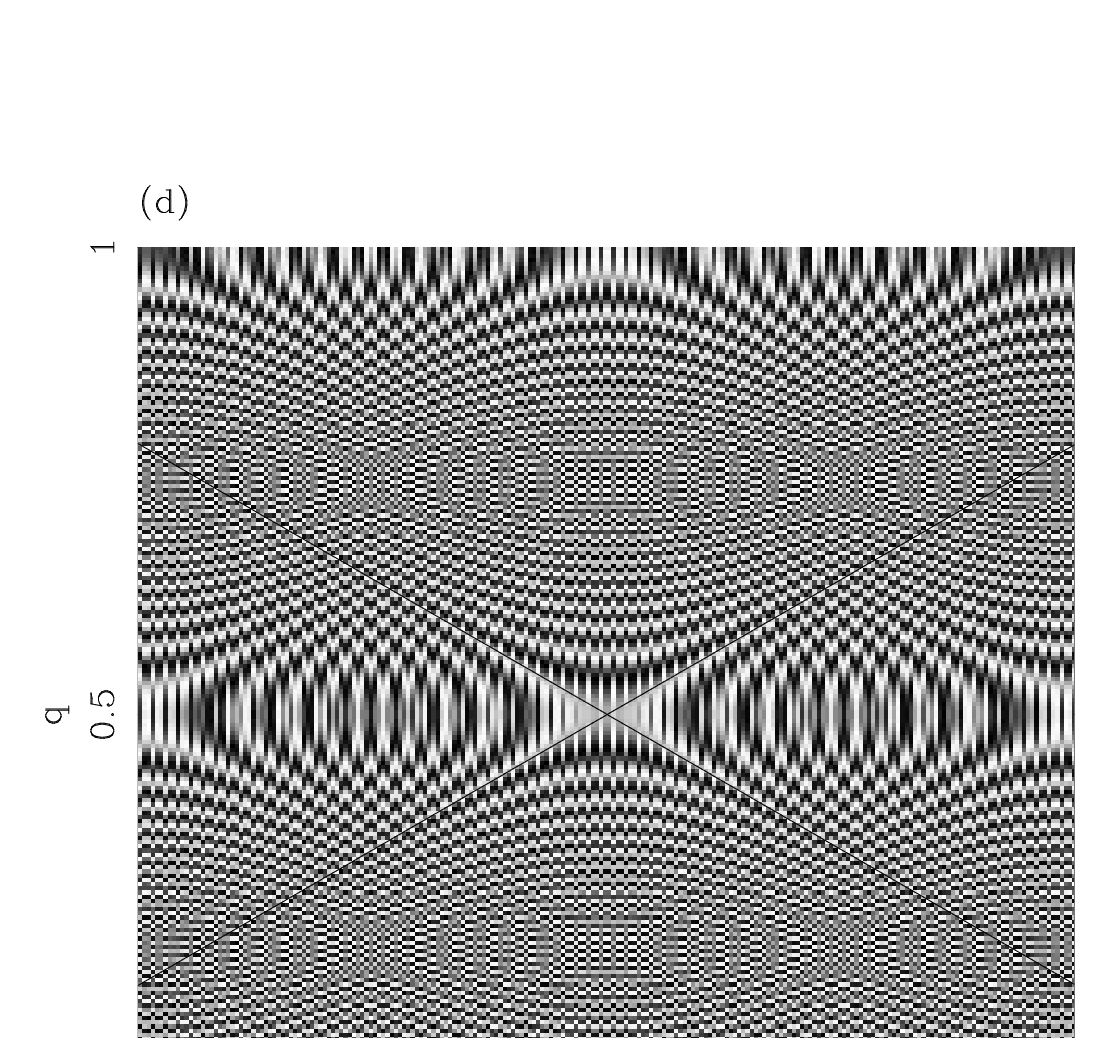}}

\end{picture}
\vspace*{1.0pc}

\caption{\footnotesize Wigner functions for the state $k=113$ with $N=223$. \\(a) individual eigenfunction, \\
(b) Spectral Wigner Function $W(x,\theta ,\epsilon ) $ for the angle $\theta$ corresponding to the quasi-energy of the state with  $k=113$ and $\epsilon=4 \Delta$, where $\Delta=2\pi /N$ is the mean level spacing of the system. \\
(c) Idem  (b) but with $\epsilon=20 \Delta$,\\
(d) Idem (b) but with $\epsilon=40 \Delta$, the black strait lines indicate the  stable and unstable manyfolds.}

\label{fig.1}
\end{figure}

%\newpage
\begin{figure}[h]
\setlength{\unitlength}{1cm}
\begin{picture}(0,16)(0,0)
\put(0.0,8.0){\epsfxsize=8cm\epsfbox[0 0 550 550]{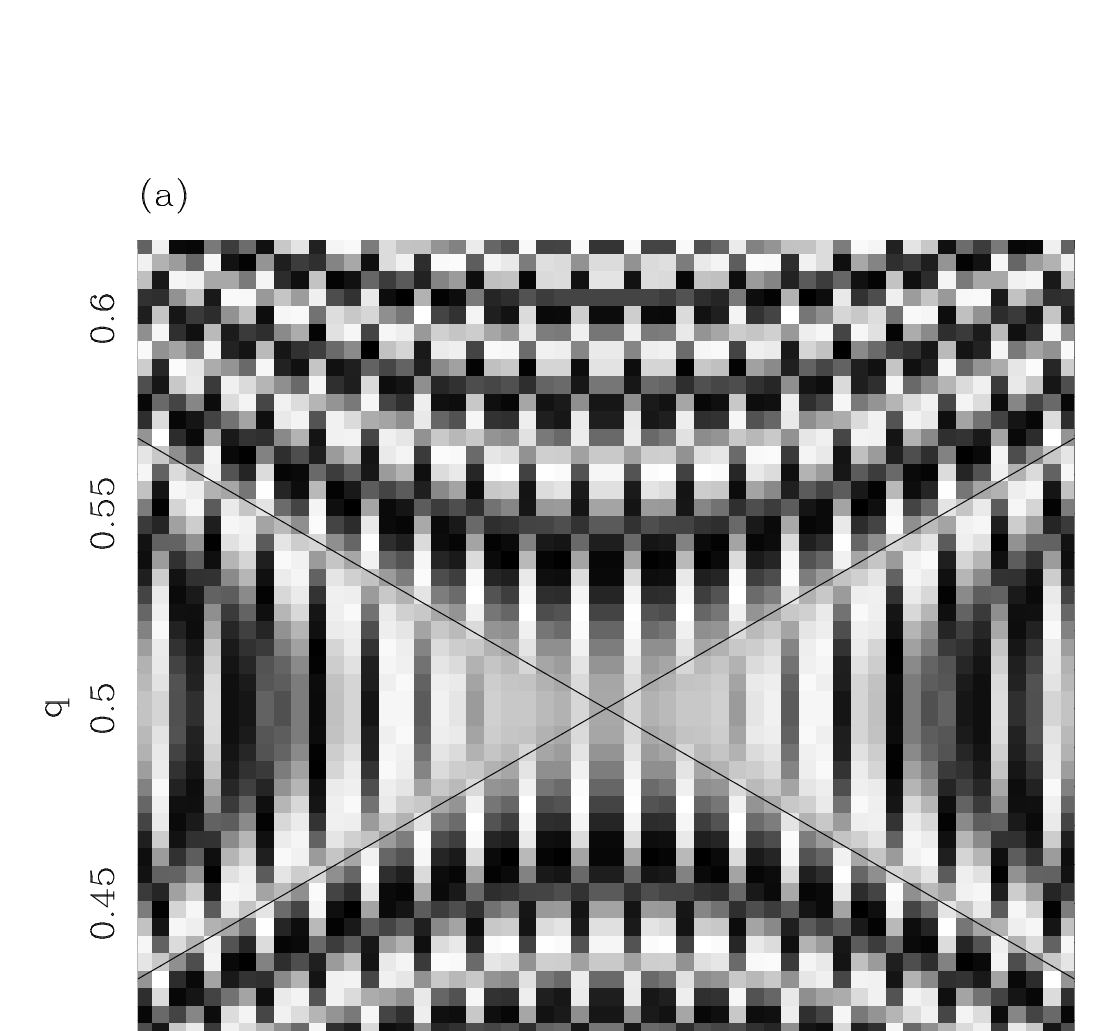}}
\put(8.0,8.0){\epsfxsize=8cm\epsfbox[0 0 550 550]{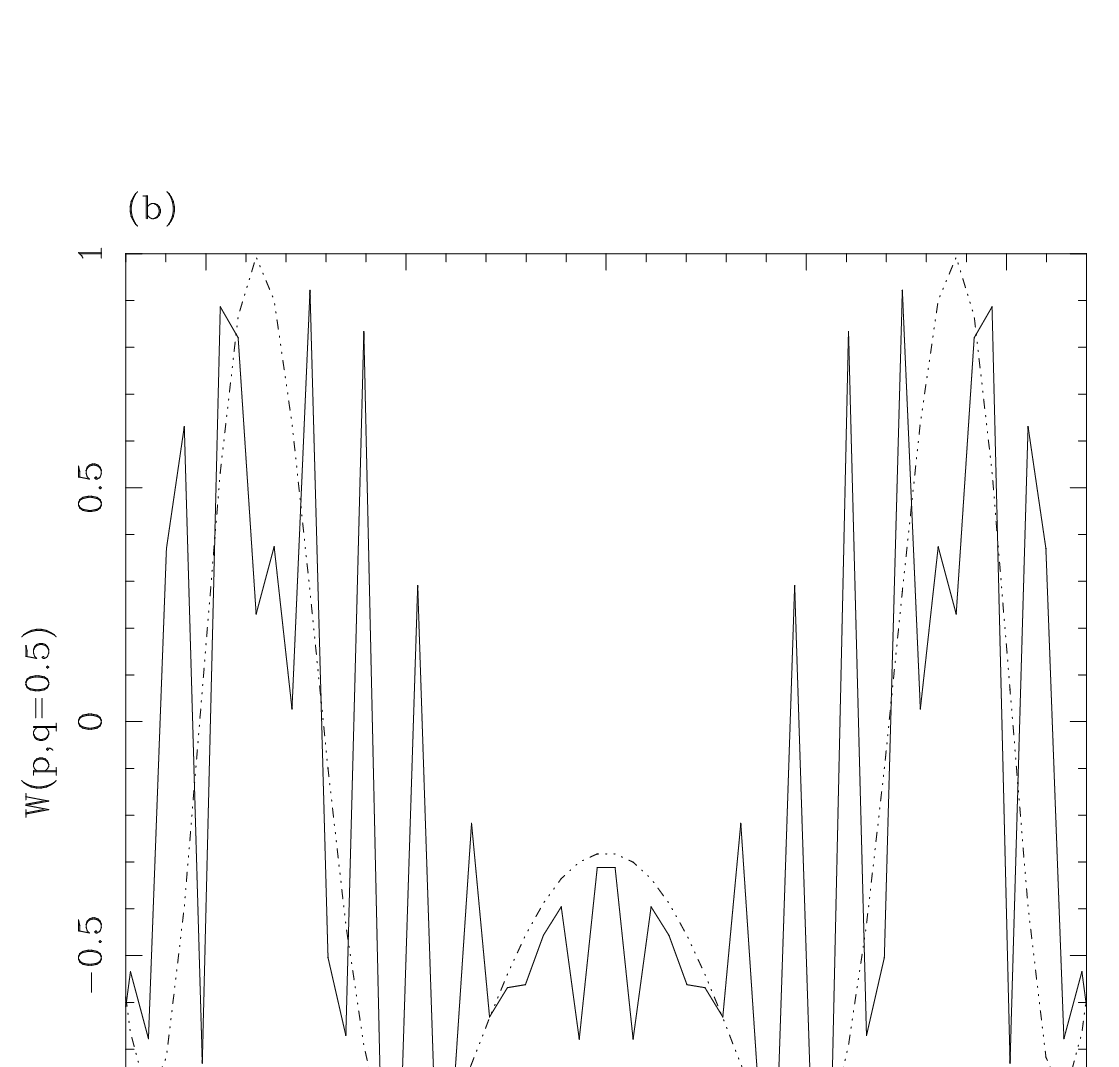}}
\put(0.0,0.0){\epsfxsize=8cm\epsfbox[0 0 550 550]{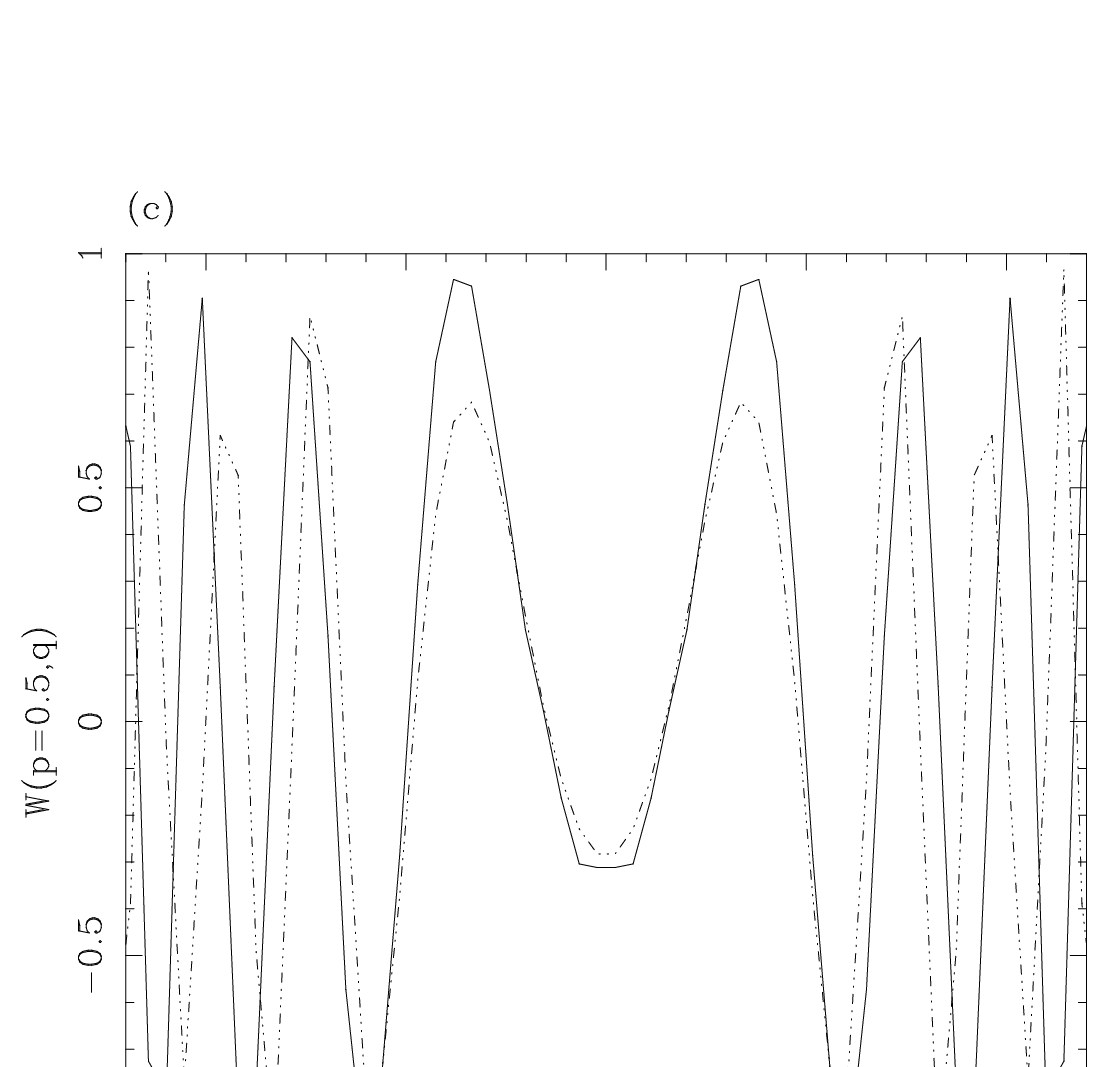}}
\put(8.0,0.0){\epsfxsize=8cm\epsfbox[0 0 550 550]{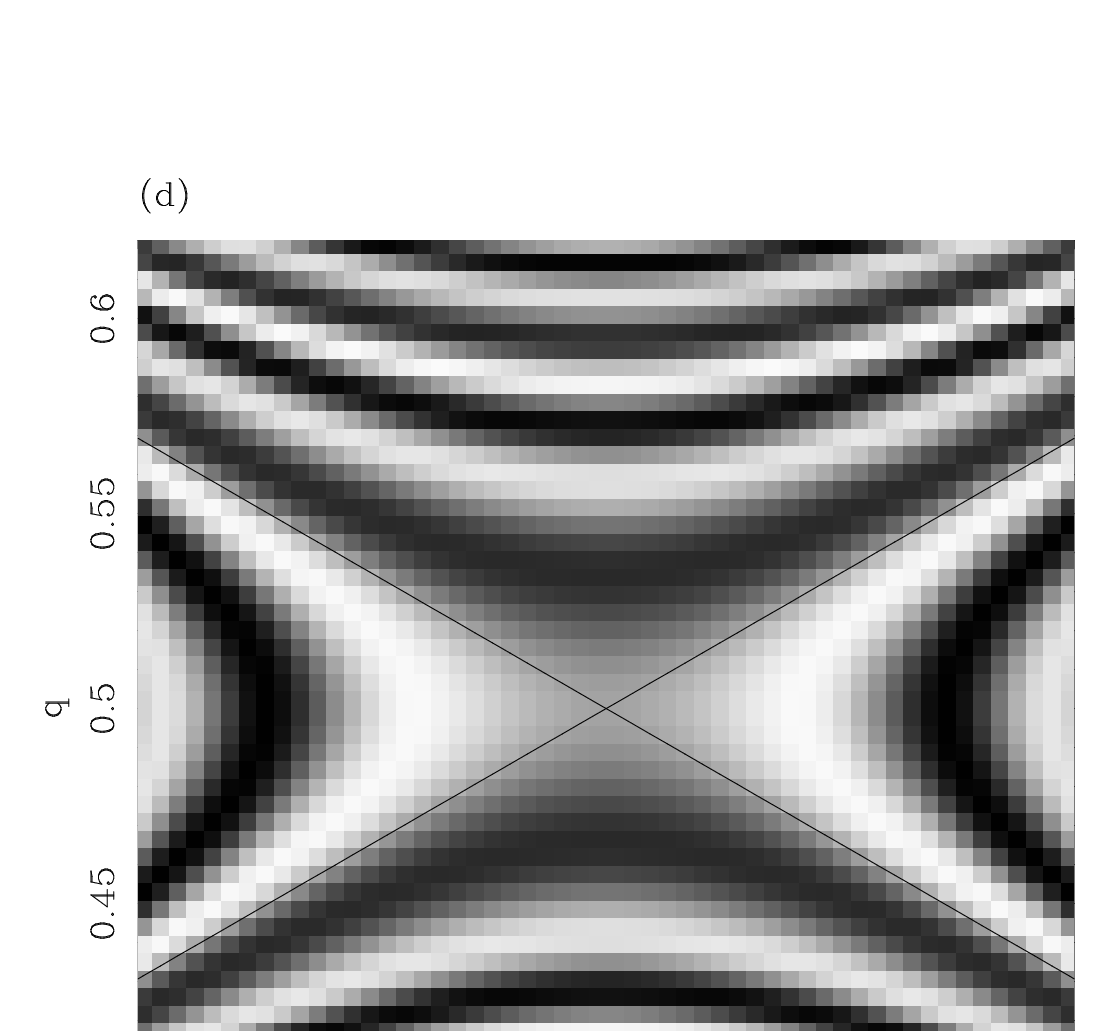}}

\end{picture}
%\centerline {\epsfxsize=17cm \epsffile{fig2.ps} }
%\centerline {\epsfxsize=8cm \epsffile{fig2.ps} }

%\centerline {\epsfxsize=6in  \epsffile{fig2.ps} }

\caption{\footnotesize  detail of the spectral Wigner function with $N=223$ near the fixed point $  [(\frac12,\frac12)]$ for an energy window of width  $\epsilon=40 \Delta$ centered at the energy of the state  $k=113$.\\
(a) Exact result for the cat map.\\
(d) Semiclassical approximation $W^j_{SC}(x,\theta ,\epsilon )$.\\
(b) We compare sections of the exact and the semiclassical Wigner functions for  $q=0.5$ (horizontal section).\\
(c) Idem (b) but for $p=0.5$ (vertical section).}

\label{fig.2}
\end{figure}

\begin{figure}[h]
\setlength{\unitlength}{1cm}
%\begin{picture}(0,6)(0,0)
%\put(0.1,0){\epsfxsize=8cm \epsfbox[83 290 512 552]{fig1.ps} }
%\centerline {\epsfxsize=17cm \epsffile{fig1.ps} }
\begin{picture}(0,16)(0,0)
\put(0.0,8.0){\epsfxsize=8cm\epsfbox[0 0 550 550]{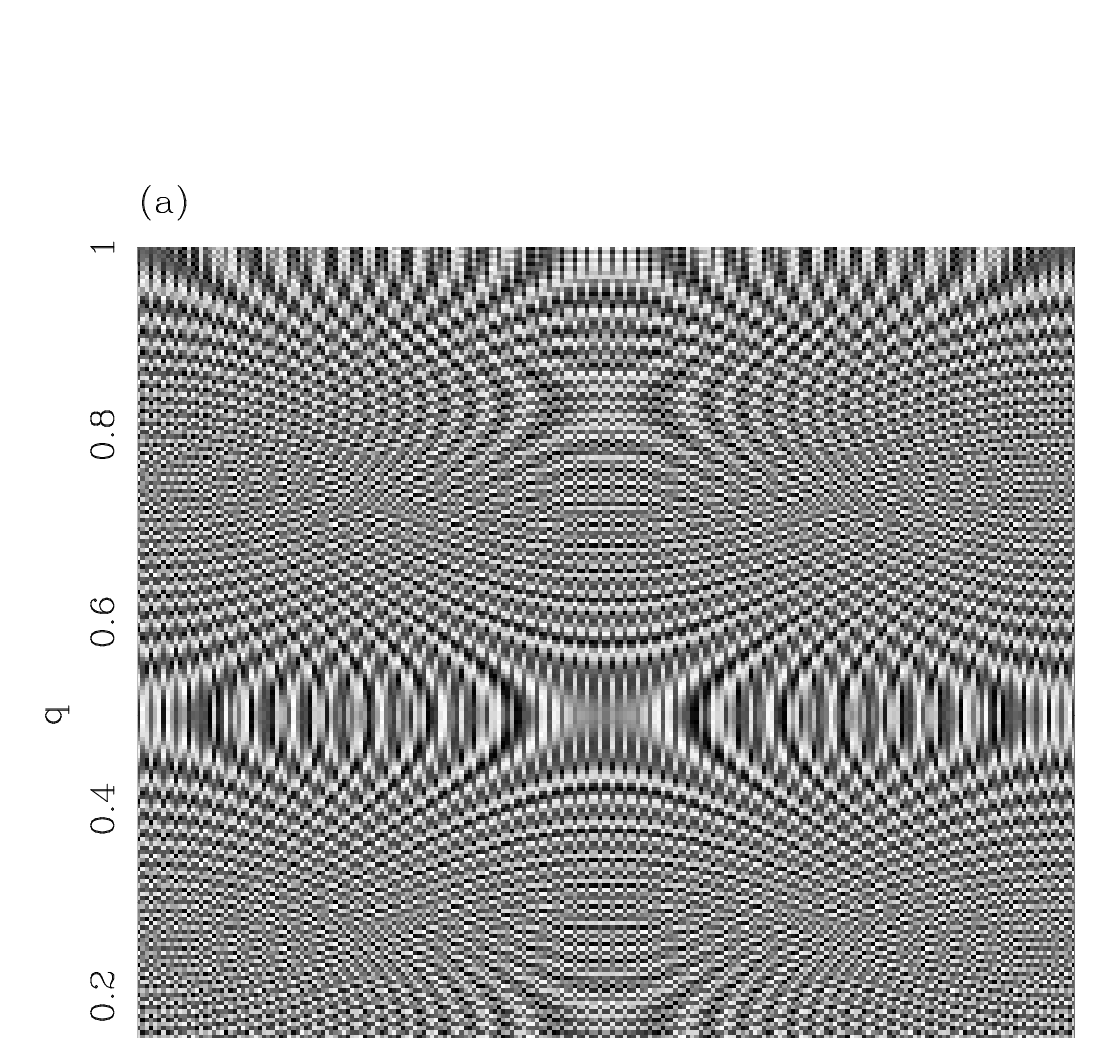}}
\put(8.0,8.0){\epsfxsize=8cm\epsfbox[0 0 550 550]{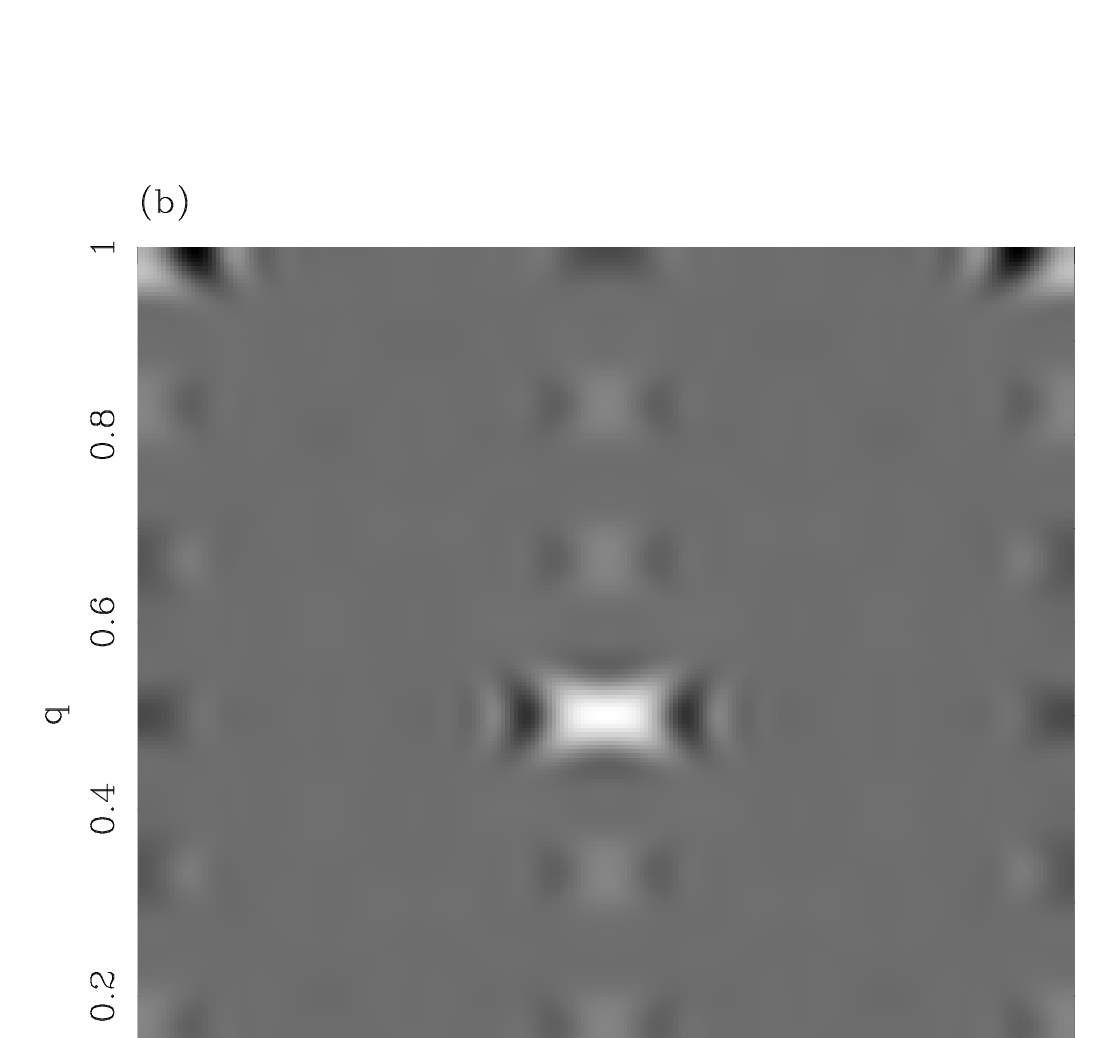}}
\end{picture}
%\vspace*{1.0pc}

\caption{\footnotesize (a) Spectral Wigner functions for an energy window centered at the state $k=113$ which is not Bohr-quantized with $\epsilon=20 \Delta$  (b) Corresponding Spectral Husimi  Function. The relative maxima along the central vertical line correpond to periodic orbits of period two.}

\label{fig.3}
\end{figure}

%\end{multicols}

\end{document}